\documentclass[aps,pre,superscriptaddress,showpacs,floatfix,twocolumn]{revtex4-1}

\usepackage{graphicx}   
\usepackage{amssymb}
\usepackage{natbib}
\graphicspath{{Plots/}}
\begin{document}

\title{Network model of human aging: frailty limits and information measures}
\author{Spencer G. Farrell}	
\affiliation{Dept. of Physics and Atmospheric Science, Dalhousie University, Halifax, Nova Scotia, Canada B3H 4R2}
\author{Arnold B. Mitnitski}	
\affiliation{Department of Medicine, Dalhousie University, Halifax, Nova Scotia, Canada  B3H 2Y9}
\author{Kenneth Rockwood}	
\affiliation{Department of Medicine, Dalhousie University, Halifax, Nova Scotia, Canada  B3H 2Y9}
\affiliation{Division of Geriatric Medicine, Dalhousie University, Halifax, Nova Scotia, Canada B3H 2E1}
\author{Andrew D. Rutenberg}	
\affiliation{Dept. of Physics and Atmospheric Science, Dalhousie University, Halifax, Nova Scotia, Canada B3H 4R2}

\date{\today}
\begin{abstract} 
Aging is associated with the accumulation of damage throughout a persons life.  Individual health can be assessed by the Frailty Index (FI). The FI is calculated simply as the proportion $f$ of accumulated age related deficits relative to the total, leading to a theoretical maximum of $f \leq 1$. Observational studies have generally reported a much more stringent bound, with $f \leq f_{max}  <1$.  The value of $f_{max}$ in observational studies appears to be non-universal, but $f_{max} \approx 0.7$ is often reported. A previously developed network model of individual aging was unable to recover $f_{max}<1$ while retaining the other observed phenomenology of increasing $f$ and mortality rates with age.  We have developed a computationally accelerated network model that also allows us to tune the scale-free network exponent $\alpha$.  The network exponent $\alpha$ significantly affects the growth of mortality rates with age. However, we are only able to recover $f_{max}$ by also introducing a deficit sensitivity parameter $1-q$, which is equivalent to a false-negative rate $q$. Our value of $q=0.3$ is comparable to finite sensitivities of age-related deficits with respect to mortality that are often reported in the literature.  In light of non-zero $q$, we use mutual information $I$ to provide a non-parametric measure of the predictive value of the FI with respect to individual mortality.  We find that $I$ is only modestly degraded by $q<1$, and this degradation is mitigated when increasing number of deficits are included in the FI. We also find that the information spectrum, i.e. the mutual information of individual deficits vs connectivity, has an approximately power-law dependence that depends on the network exponent $\alpha$. Mutual information $I$ is therefore a useful tool for characterizing the network topology of aging populations.
\end{abstract}
\pacs{87.10.Mn, 87.10.Rt, 87.10.Vg, 87.18.-h} 
\maketitle

\section{Introduction} 

Humans above the age of $40$ experience an exponential increase in mortality rate with age, known as Gompertz's law~\cite{Gompertz, Kirkwood:2015}. We can view aging as the accumulation of damage over time~\cite{Kirkwood:2005}. However, individual health status  increasingly varies as age increases~\cite{Rockwood:2000}.  Quantitative measures of individual aging-related health that measure the accumulation of damage throughout a persons life are useful for predicting adverse outcomes in older populations such as loss of independence, hospitalization, surgical complications, and mortality~\cite{Mitnitski:2001, Rockwood:2005}.   

The Frailty Index (FI) is a quantitative age-related measure of health~\cite{Mitnitski:2001, Rockwood:2002, Mitnitski:2013, Kulminski:2007, Yashin:2008} that provides a  score $f \in [0,1]$.  To determine $f$,  distinct deficits (aspects of age-related health) are assessed clinically and assigned values of $0$ for the absence of a deficit (healthy) or $1$ for the presence of a deficit (damaged). Each deficit is weighted equally, and $f$ is calculated as the fraction of damaged deficits, typically using $N \approx 30-40$  deficits~\cite{Searle:2008}. Arithmetic provides fundamental limits of $0 \leq f \leq 1$.   

A large body of clinical and epidemiological work has shown that the FI correlates strongly with mortality~\cite{Rockwood:2002, Mitnitski:2005, Kulminski:2007}, and increases nonlinearly with age~\cite{Kulminski:2011}.  In older people, the FI also correlates with postoperative complications~\cite{Makary:2010, Partridge:2011}, risk of hospitalization, and risk of dependence~\cite{Jotheeswaran:2015}. Distributions of the FI broaden with age, capturing the increasing variation in individual health~\cite{Gu:2009, Mitnitski:2013}.  A broad range of possible age-related deficits can be used to calculate $f$~\cite{Rockwood:2006ab}, indicating that the FI is robust to the details. Intriguingly, there is an observed upper limit $f \leq f_{max} \approx 0.7-0.8$ that is significantly below the arithmetic limit~\cite{Searle:2008, Gu:2009, Mitnitski:2013, Bennett:2013, Hubbard:2015, Armstrong:2015}. Nevertheless, the precise value of $f_{max}$, as assessed by the $99$th percentile value of $f$ in a cohort of frail elderly, is not universal. For example, $f_{max} \approx 0.5$ has been observed in a large UK study using electronic health records \cite{Clegg:2016} and in the Study on Global AGEing and Adult Health (SAGE) \cite{Harttgen:2013}, while $f_{max} \approx 0.3$ from GP records in the Netherlands \cite{Drubbel:2013}.

To address a possible origin of $f_{max}$, we build upon a recent stochastic network model of aging by Taneja {\em et al.}~\cite{Taneja:2016}. In that model, which used a scale-free network topology, nodes correspond to individual deficits. Local damage and repair rates depend on the local state of the network; damage of a particular node is faster and repair slower as its connected neighbours become more damaged. The interactions between deficits capture some of the complex nature of interacting health conditions. Mortality results in the damage of the most highly connected nodes, while the FI is assessed from highly connected nodes that are distinct from the mortality nodes.  This model qualitatively captures the Gompertz-like exponential growth of mortality rate at later ages, the evolution of the FI with age, and the broadening of frailty distributions with age~\cite{Taneja:2016}. The network model of Taneja {\em et al}~\cite{Taneja:2016} has no explicit time-dependence in damage or repair rates, or in its mortality condition.  It represents aging as an autonomous and non-adaptive accumulation of health deficits, the generally accepted view, and stands in contrast to picture of programmed aging \cite{Vijg:2016}. Nevertheless, the Taneja model could only recover observed values of the FI limit $f_{max}$ by significantly overestimating mortality in younger adults. While an underestimation of mortality could be corrected by  mortality processes exogenous to the model, an {\em overestimation} cannot be and so represents a significant open issue.

We are aware of three hypotheses for the origin of the FI limit. First: that $f_{max}$ arises naturally in the aging process through a large effective repair rate that prevents extremes of damage or a large mortality rate that makes it extremely unlikely to live beyond $f_{max}$. In terms of a quantitative model, this amounts to a parameter choice.  Taneja {\em et al} could not find a working parameterization \cite{Taneja:2016}. Furthermore, the observed non-universality of $f_{max}$ between similar populations, as noted above, argues against any such intrinsic origin.  Second: that mortality occurs at $f_{max}$.  Such a threshold networked model has been developed to explore non-human mortality~\cite{Vural:2014}, though it was not used to  explore the FI phenomenology. Thresholded mortality does not explain the non-universality of $f_{max}$, but does raise an interesting question of programmed mortality (as opposed to programmed aging).  The third hypothesis that we propose is novel: that the apparent $f_{max}$ observed in the clinical data reflects limited sensitivity of clinical diagnosis of deficits. Such limited sensitivity is intrinsic to any clinical assessment due to fundamental tradeoffs with respect to specificity, and can be characterized with receiver operating characteristic (ROC) curves~\cite{Metz:1978, Zweig:1993}. This third hypothesis provides a simple explanation of a non-universal FI limit, since different studies include different deficits and will have different sensitivities. Furthermore, we could reconcile the third hypothesis (but not the first two) with observed aging phenomenology using our improved network model.

The significance of the FI is due to its predictive capacity for health outcomes. This has been assessed parametrically vis-a-vis mortality, through a proportional  \cite{Mitnitski:2005, Kulminski:2007} or quadratic \cite{Yashin:2008} hazards model. Non-parametric assessment has been mostly qualitative, through separation of survival curves that are stratified by the FI  -- see e.g. \cite{Clegg:2016}. Information theory provides a quantitative and non-parametric measure, and has been proposed for mortality statistics \cite{Steinsaltz:2012, Blokh:2016}.  

Information entropy or Shannon entropy $S(A)$~\cite{Shannon, Cover} is a quantitative measure of uncertainty in a random variable $A$ with probability distribution $p(a)$. For a discrete (binned) distribution, then $S(A) = -\sum_a p(a)\ln{p(a)}$.  Entropies of conditional death age distributions allow us to quantify the information added to the unconditioned distribution. If $S(A|t)$ is the uncertainty remaining about the death age $A$ given that the person has survived to specific age $t$, the difference $I(A;t) \equiv S(A) - S(A|t)$ is the reduction of uncertainty by knowing the age $t$ --- and is the information gained. Similarly, the  information gained by knowing the FI at a given age $t$ will be $I(A;f|t) \equiv S(A|t) - S(A|f, t)$. If we average over all FI values given the specific age, the average information gained by knowing a persons FI at a given age compared to just knowing their age is $I(A;F |t) \equiv S(A|t) - S(A|F, t)$, where the capital $F$ indicates an average over values of $f$. This is called the mutual information between the death-age and the FI at a given age $t$. 

We use mutual information to non-parametrically assess the predictive value of our model FI with respect to the death-age distribution. We characterize how much information  knowing a persons age adds;  how much information the FI  adds; and how much information individual deficits provide.  We are able to address how the predictive information of the FI, with respect to mortality, is degraded in the face of sensitivity errors. We find, at the levels called for by the observational $f_{max}$, that the information loss is not substantial. We also find that information measures are sensitive to the topology, and so should offer insight into the relations between clinical deficits.

\section{Model and Analysis} 

Our model is a simplified, extended, and accelerated adaptation of the model of Taneja  {\em et al.}~\cite{Taneja:2016}. Our model differs by including a tuneable rather than fixed scale-free exponent ($\alpha$), by using exponential (but empirically similar) damage and repair rate dependence on the $f_i$ rather than Kramer's rates from an asymmetric double-well potential, by using two mortality nodes that must be simultaneously damaged for mortality rather than one, and by significantly improving the numerical implementation ($\approx 10^4$ speedup) to allow many more nodes and many more individuals to be simulated

Each individual is represented by a randomly generated scale-free network consisting of $N$ nodes, where each node $i \in \{1, 2, \ldots, N \}$ corresponds to a deficit that can take on binary values $d_i = 0$ or $d_i = 1$ for healthy or damaged, respectively. Connections are undirected, and all deficits are initially undamaged at $t=0$. When nodes damage or repair, connections are unaffected. We generate a scale-free network~\cite{Albert:2002} with degree distribution $P(k) \sim k^{-\alpha}$, where $k$ is the degree of a node, using the Barab\'{a}si-Albert preferential attachment model~\cite{BA1999}, using a linear shift to tune the exponent $\alpha$~\cite{Krapivsky:2000}. This allows us to independently adjust both the exponent $\alpha$ and the average degree $\langle k \rangle$.  The two most highly connected nodes are  mortality nodes, and when both are in the damaged state, mortality occurs. [The effect of different numbers of mortality nodes has been explored previously~\cite{Taneja:2016}.] Because of the scale-free character of the network, mortality nodes are much more connected than most other nodes in the network. This follows our intuition that mortality is impacted by many factors.

For the $i$th node, healthy deficits damage at rate $\Gamma_+ = \Gamma_0 \exp{(f_i \gamma_+)}$ and damaged deficits repair at rate $\Gamma_- = (\Gamma_0/R) \exp{(-f_i \gamma_-)}$. The damage and repair rates depend on the average deficit value of all connected nodes, $f_i$. This local frailty $f_i$ is a dynamical variable, since it changes along with the connected deficits. The other parameters, $\gamma_+$, $\gamma_-$, $\Gamma_0$ and $R$, are all time-independent and the same for all nodes --- including mortality nodes. Transitions are implemented exactly using Gillespie's stochastic simulation algorithm (SSA)~\cite{Gillespie:1977}, also known as kinetic Monte Carlo (kMC), using a binary tree method to efficiently identify which deficit changes~\cite{Gibson:2000}.

The FI is calculated as the average deficit value, $f = \sum_i^n d_i/n$ over the $n$ most connected network nodes that are not mortality nodes. These ``frailty nodes'' typically represent a small fraction of all deficits, and are diagnostic. Since frailty nodes are highly connected, they should provide a good measure of the average health of the network -- just as the clinical FI provides a good measure of human health.

Our model results are based on a simulated population of $10^7$  individuals and $N = 10^4$ (number of network nodes). Each individual network is stochastically evolved in time until mortality. Our default parameters are $\gamma_+ = 10.27$, $\gamma_- = 6.5$, $R = 1.5$, $n = 32$ (number of FI deficits), $\alpha = 2.27$, and $\langle k \rangle = 4$.   The only dimensional parameter is the overall damage rate, $\Gamma_0 = 0.00113$ (per year).  Parameters were chosen to give qualitative agreement with population mortality rates, the average FI trajectory, and FI distributions from observational data.    A deterministic version of our model, equivalent to a maximally-connected network, is presented in Appendix~\ref{appendix:deterministic}. In Appendix~\ref{appendix:parameters} we explore the roles of repair rates and the scale-free exponent $\alpha$.  

We implement finite sensitivity $1-q$ through a false-negative rate $q$.  False-negative rates are applied to every individual FI and have no effect on the dynamics.  For an uncorrected individual FI value of $f_0$ from $n$ deficits in the FI, there are $n_0 = f_0 n$ damaged nodes. With a false-negative rate $q$, we record only $n_q$ damaged nodes where $n_q$ is sampled from the binomial distribution $p(n_q) = { n_0 \choose n_q} (1-q)^{n_q} q^{n_0-n_q}$.  We then use $f=n_q/n$ as the corrected individual FI.  On average, we will obtain $\langle n_q \rangle = (1-q) n_0$. We use a default false-negative rate of $q=0.3$, unless otherwise noted.  

Information entropies are estimated directly from a list of $M$ ordered individual death ages $\{a_i\}$~\cite{Vasicek:1976, Dudewicz:1981, vanEs:1992, Bierlant:2001, Miller:2003}. The entropy is calculated using
\begin{equation} \label{list}
	S(A) = \frac{1}{M - m}\sum\limits_{i = 1}^{M - m} \ln{(a_{i+m} - a_i)} - \psi(m) + \psi(M + 1),
\end{equation}
where $\psi$ is the digamma function~\cite{vanEs:1992, Miller:2003}. We require that $M\gg m\gg 1$, and we use $m = \sqrt{M}$ to reduce noise in the entropy calculation~\cite{Bierlant:2001, Miller:2003}. 

To calculate conditional entropies averaged over the FI, $S(A|F, t)$, death age lists $\{a_i\}$ are binned by current age and FI, $p(A|f, t)$. Then using frailty distributions $p(f|t)$, entropy is calculated by averaging over the FI: $S(A|F, t) = \sum_f P(f|t) S(A|f,t)$. This allows us to calculate mutual information, $I(A;F |t) = S(A|t) - S(A|F, t)$. We are also interested in the information provided by specific values of the FI; the specific mutual information. To calculate the specific mutual information $I(A;f|t) = S(A|t) - S(A|f,t)$, we do not average over the FI. In this notation, capital letters denote values that are averaged over, and lower case letters indicate specific values of the variable. Bin widths of $0.01$ are used to average over the FI, and of $1$ year for death age distributions. 

\section{Results} 

\begin{figure} 
  \begin{minipage}[t]{0.45\textwidth}
    \includegraphics[width=.9\textwidth]{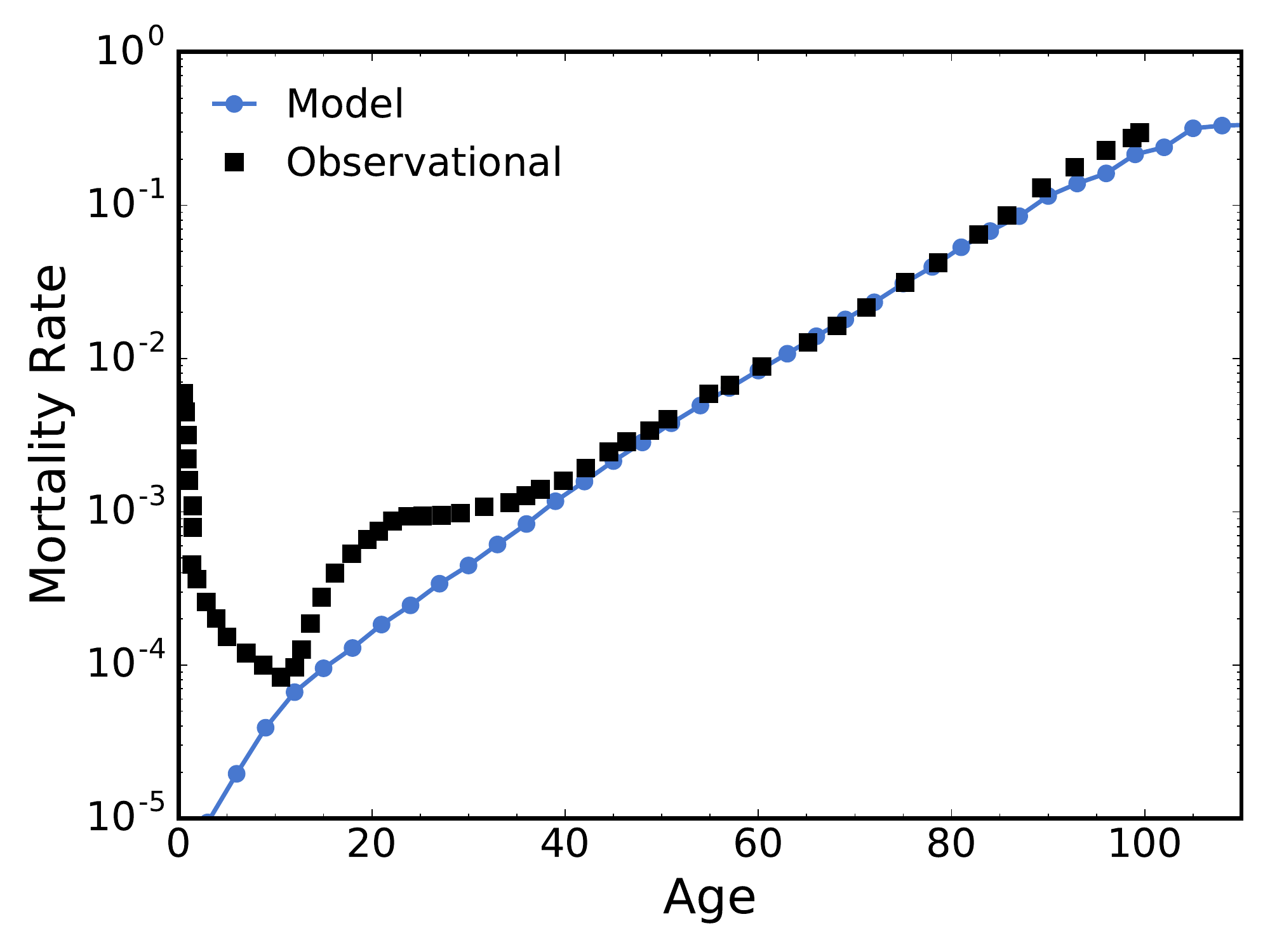}
    \caption{Mortality rate vs age for the model (blue circles) and US population mortality statistics (black squares). Default parameters were used for the model, including $q=0.3$. Mortality statistics are from~\cite{Arias:2014}. All ages in this and subsequent figures are in years. Mortality rates are per year.}
    \label{Mortality}
  \end{minipage}
\end{figure}

\begin{figure} 
  \begin{minipage}[t]{0.45\textwidth}
    \includegraphics[width=.9\textwidth]{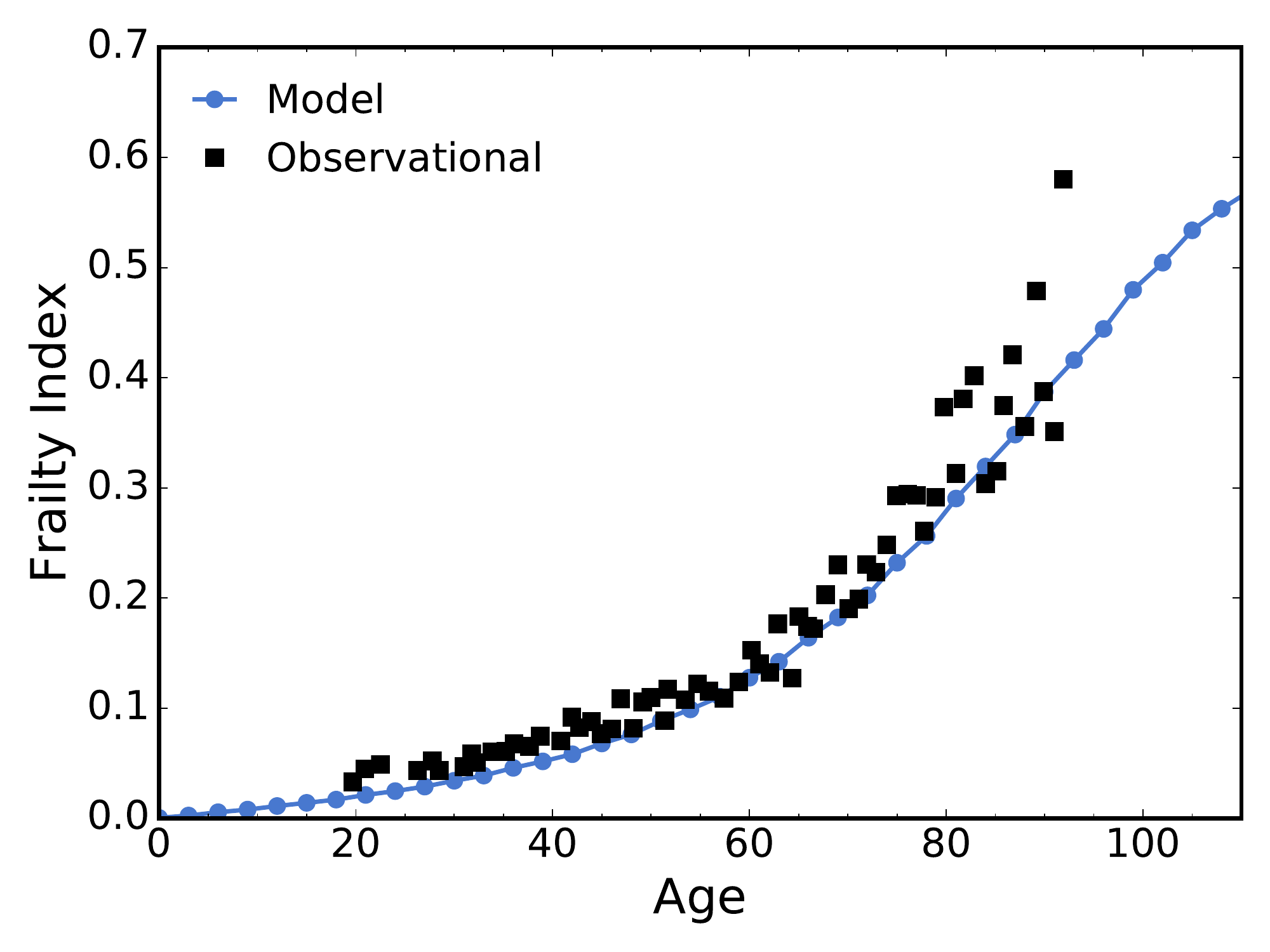}
    \caption{Average FI vs age for the model (blue circles) and observational FI data (black squares). Default parameters were used for the model, including $q=0.3$. Observational data is from~\cite{Mitnitski:2013}.}
    \label{Frailtyq03}
  \end{minipage}
\end{figure}

Fig.~\ref{Mortality} shows the model mortality rate vs age in blue with United States mortality rate statistics~\cite{Arias:2014} in black. Fig.~\ref{Frailtyq03} shows the model average FI vs age in blue with FI data from the Canadian National Public Health Survey (NPHS)~\cite{Mitnitski:2013} in black. For ages above 40, which is the focus of our model, we obtain good agreement for the mortality rate vs age and for the average FI vs age. The agreement of the age-dependent mortality with our model is better than, and of the FI phenomenology with our model is similar to, the agreement that Taneja {\em et al.} \cite{Taneja:2016} could obtain. This shows that including our default false-negative rate ($q=0.3$) and other model adjustments can be accommodated by variations of the  model parameters. 

\subsection{FI Limit}

\begin{figure} 
  \includegraphics[width=.39 \textwidth]{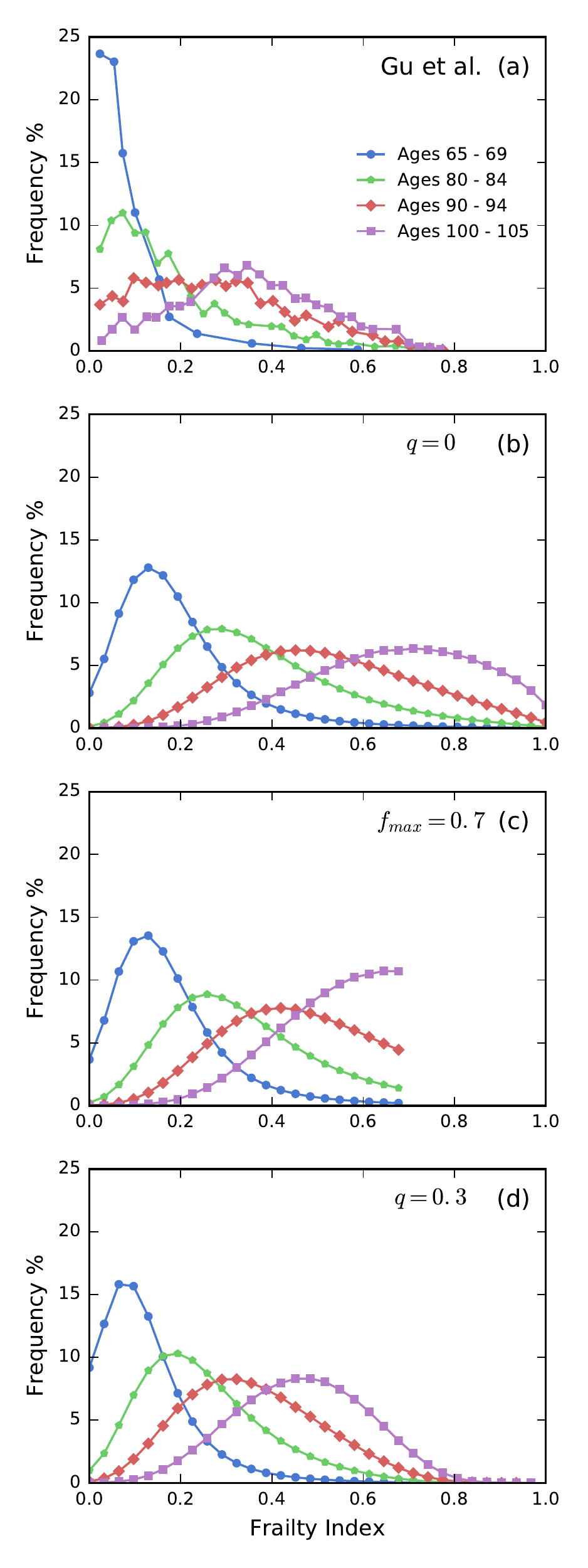}
  \caption{Distributions of the FI in a given age range, $p(f|t)$. Age ranges are indicated by the legend. {(a)} Chinese population observational data from Gu {\em et al.}~\cite{Gu:2009}. Note the FI limit around $0.8$. {(b)} Model distributions without  a false negative rate, i.e., $q = 0$. This is the first hypothesis for the FI limit. We observe $f_{max}=1$. {(c)} Model distributions with additional mortality imposed at $f =f_{max}= 0.7$ but with $q=0$. This is the second hypothesis for the FI limit. There is a discontinuous cut-off in the FI distributions. {(d)} Model distributions with our default false-negative rate of $q = 0.3$. This is our third hypothesis for the FI limit. We find $f_{max}=0.78$ at the 99th percentile of the population of 100-105 year olds.}
\label{FIDistributions}
\end{figure}

Fig.~\ref{FIDistributions}(a) shows FI distributions for selected age ranges of Chinese population data from Gu {\em et al.}~\cite{Gu:2009}. The limit in FI is seen as a maximum value around 0.7 - 0.8. Fig.~\ref{FIDistributions}(b) shows FI distributions from our model using default parameterization but with $q=0$ (no false-negatives). While we are able to capture the time-dependence of the mortality and FI with $q=0$ (data not shown), and we were able to capture the increasing variation in individual health with age seen in the FI distributions, we were unable to capture the FI limit at the same time.   We found the same limitation in a deterministic formulation of our model (see Appendix~A) that could rapidly explore the model parameters.  Our inability to find parameters that recover $f_{max}$  agrees what was reported by Taneja {\em et al.} \cite{Taneja:2016}, despite our now being able to additionally vary the scale-free-exponent $\alpha$. 

We also examined the second hypothesis, by adding a mortality condition whenever $f>f_{max}=0.7$ that is in addition to our standard two-node mortality condition with $q=0$.  Fig.~\ref{FIDistributions}(c) shows the FI distribution from this hybrid mortality model with an explicit FI threshold.  As expected $f<f_{max}$ is reproduced, and also the mortality and FI evolution (data not shown). However, a strong discontinuity is seen in the FI distribution at $f_{max}$ for older age ranges. This is not observed in the population data of Fig.~\ref{FIDistributions}(a). Correspondingly, a peak in the mortality vs $f$ is observed at $f_{max}$ that is not observed in the population data \cite{Mitnitski:2006} (data not shown). While one could consider spreading the mortality over a range of $f$ to soften these non-analyticities, the observed non-universality of the observed $f_{max}$ would remain difficult to reconcile with this intrinsic mortality mechanism.

\begin{figure*} 
  \includegraphics[width=\textwidth]{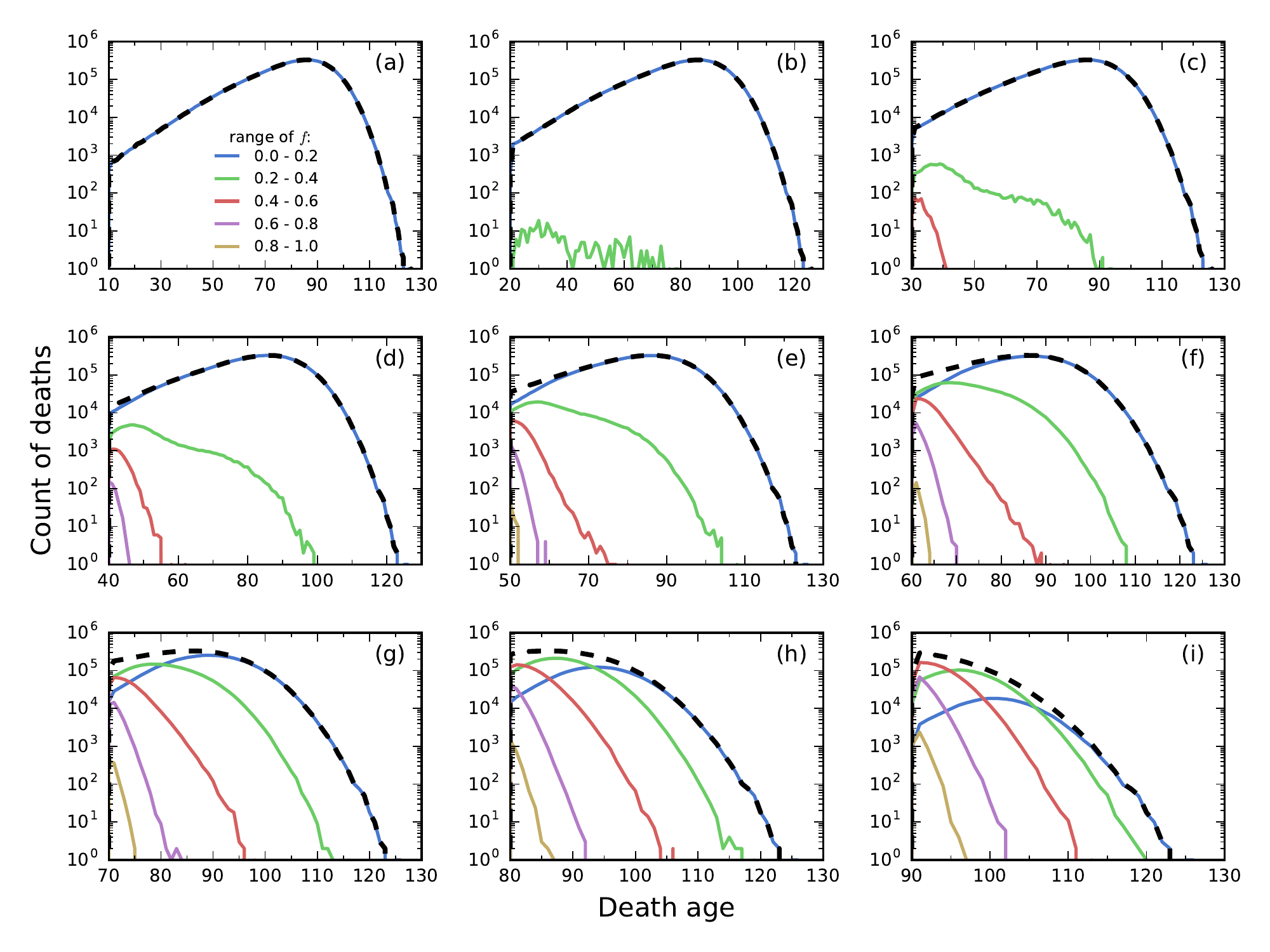}
  \caption{Unnormalized distributions of model death ages $a$ conditional on being alive at $t$, $p(a|t)$, are shown as thicker dashed black lines. These are death age or age-at-death distributions. From (a)-(i), the initial age $t$ increases from $10-90$ years, as indicated by the earliest age shown. The original population is $10^7$ model individuals. The coloured lines show the death ages $p(a|f,t)$  conditioned by the FI ranges, as indicated by the legend in (a).  As initial age $t$ increases, more FI ranges are populated. All data is binned in one year increments.  Default parameters are used, including $q=0.3$.}
\label{ConditionalDistributions}
\end{figure*}

Fig.~\ref{FIDistributions}(d) shows the result using a false negative rate $q=0.3$ (our default parameterization), our third hypothesis for the origins of the FI limit. This is imposed on the analysis of the FI only, and has no effect on mortality. We see that a FI limit is recovered, with $f_{max}=0.78$ at the 99th percentile. We have already seen that the Gompertz law, Fig.~\ref{Mortality}, and the non-linear increase of the FI with age,  Fig.~\ref{Frailtyq03}, are recovered with $q=0.3$.  This appears to be the simplest approach that works within the context of our model. It has the advantage of naturally explaining the non-universality of $f_{max}$ in terms of the non-universality of something extrinsic to aging and mortality -- namely the  sensitivity (with respect to mortality) of the deficits used in a given study. Since finite sensitivity (i.e. $q>0$) is typically where clinical assessment operates \cite{Zweig:1993}, we view this as a parsimonious and successful extension to our initial model. 

\subsection{Mutual information of the FI and mortality}

Fig.~\ref{ConditionalDistributions} shows unnormalized death age distributions, with number of deaths in $1$ year bins from an initial population of $10^7$ model individuals. Each subfigure corresponds to the subpopulation alive at the earliest age shown, i.e. $10-90$ years for (a)-(i), respectively. The thicker dashed black lines show $p(a|t)$, the death age distribution conditioned on that earliest age $t$, i.e. the number of people that die at each age $a$ given that they have already lived to age $t$. The colored lines, as indicated by the legend in (a), show death age distributions conditioned on both age and the FI value $f$, i.e. $p(a|f, t)$. This is the number of people that die at each age $a$, given they were alive at age $t$ with $f$ in the indicated range. As the initial age $t$ increases, more of the population is found at higher FI ranges.   We see that cohorts with lower FI die later, while cohorts with larger FI die earlier.  Summing over all of the FI cohorts returns the distributions conditioned on age alone, i.e. $p(a|t) = \sum_f p(a|t,f)$.

We see from Fig.~\ref{ConditionalDistributions} that increasing the initial age $t$  narrows the death-age distribution. For all but the youngest initial ages, conditioning on the FI further narrows the death-age distributions. This narrowing reflects additional predictive value due to the FI, which we can quantify with mutual information. 

\begin{figure}  
  \centering
  \includegraphics[width=0.46\textwidth]{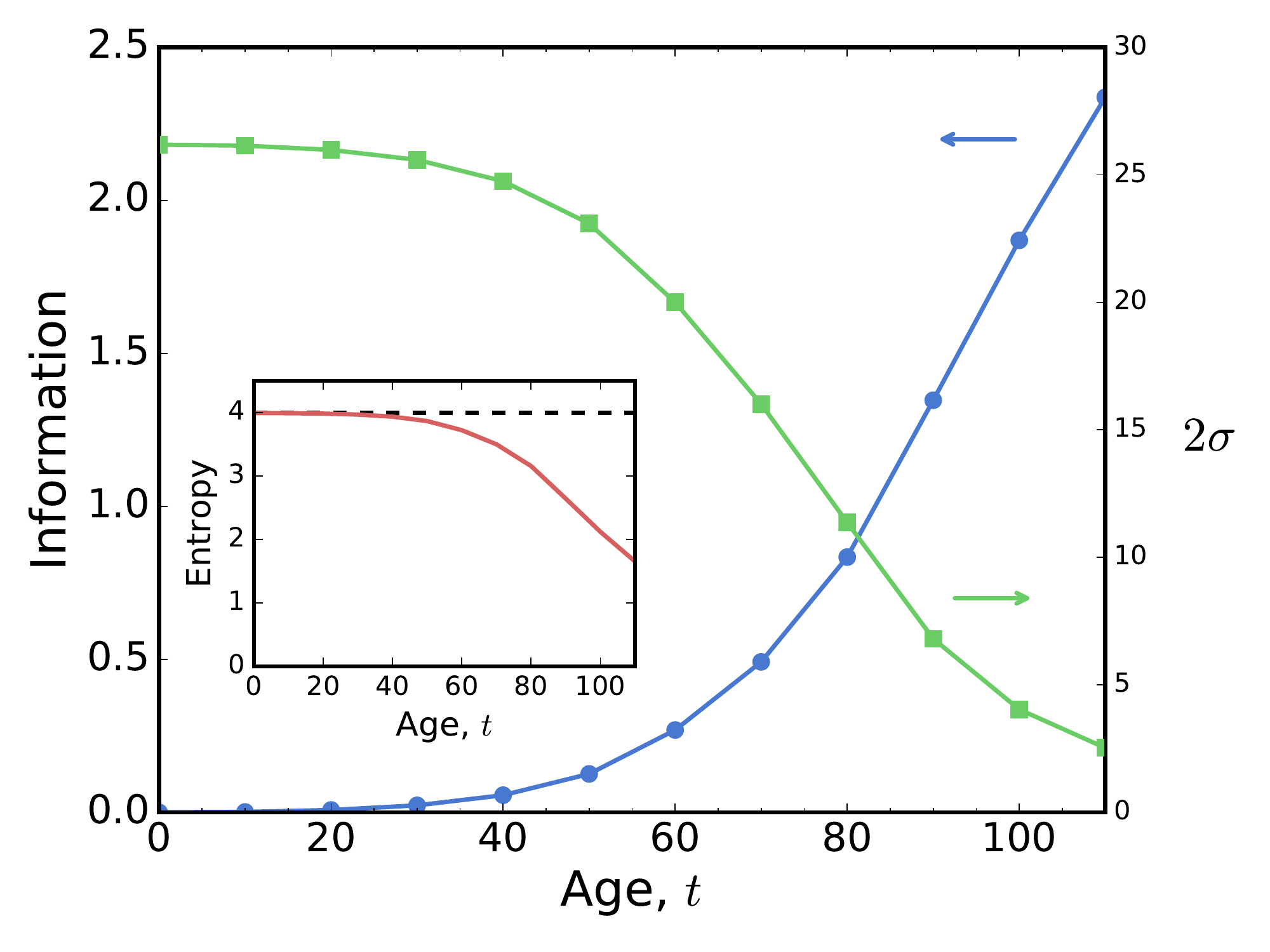}
  \caption{Information $I(A;t) = S(A) - S(A|t)$ is plotted vs age $t$ (blue points, left axis). This is the information gained about a model individual's death age by knowing their age, compared to knowing just the population distribution of death ages. The Gaussian width $2 \sigma$ (in years) that would provide the same information is also shown (green points, right axis). The inset shows $S(A|t)$ in blue, and $S(A)$ as a black dashed line.}
  \label{ModelInformationAge}
\end{figure}

Fig.~\ref{ModelInformationAge} shows the specific mutual information $I(A;t) = S(A) - S(A|t)$ of the age $t$ vs $t$ (blue points referring to the left axis). This is the information gained at a specific age  $t$ compared to having no knowledge of $t$. The inset shows constant population entropy $S(A)$ vs the entropy conditioned on survival to age $t$, $S(A|t)$. At age $0$ years old, we know only as much as we do for the whole population, so $I(A;0)=0$. As age $t$ increases, more information is known about an individual's death age, as also reflected by the narrowing of the death-age distribution with age shown in Fig.~\ref{ConditionalDistributions}(a)-(i). With the green points (referring to the right axis) we have shown the width of the Gaussian $2 \sigma$ that would give the same information. This allows us to roughly convert information to an age-range.

\begin{figure} 
  \centering
  \includegraphics[width=0.46\textwidth]{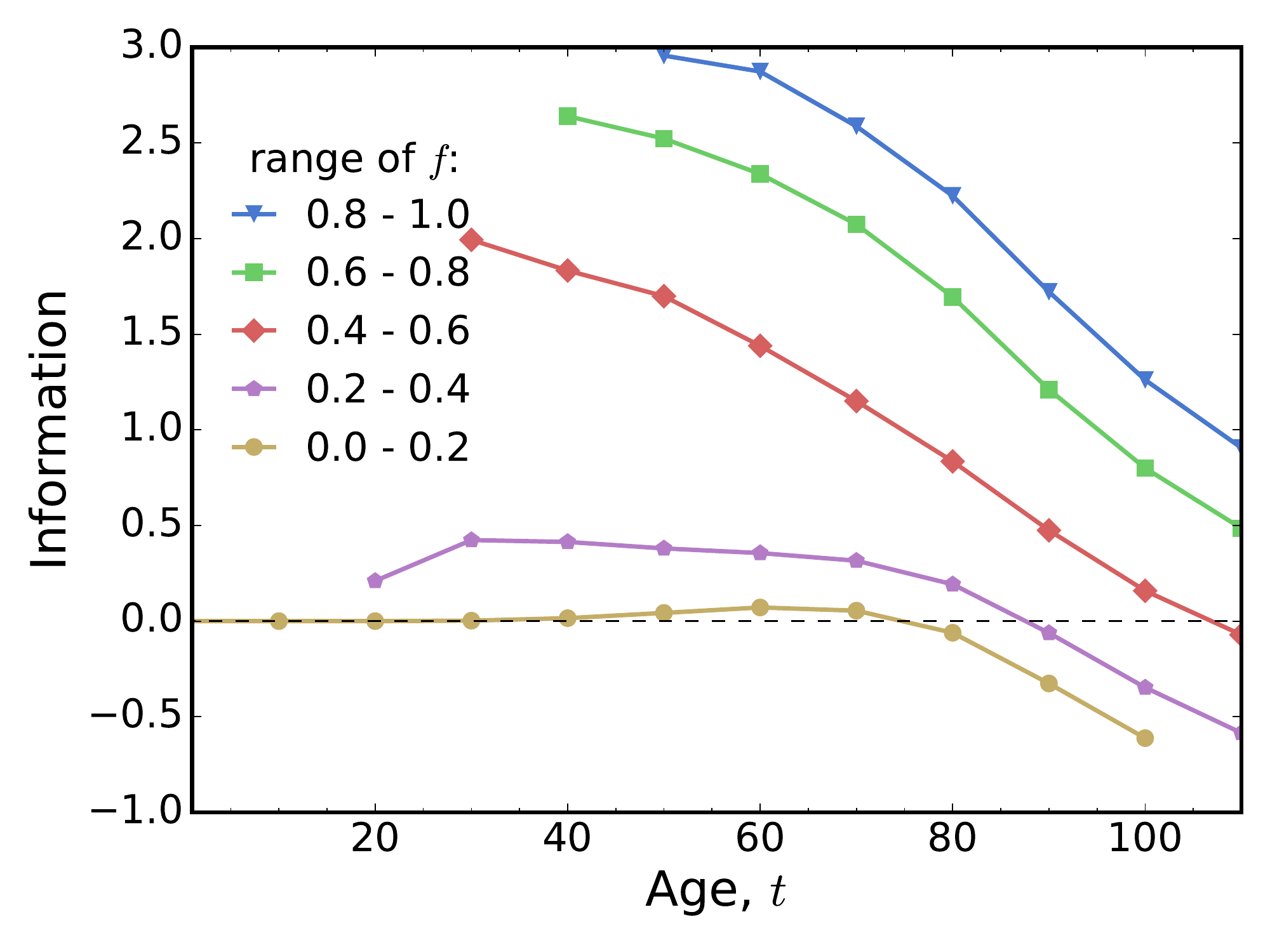}
  \caption{Specific mutual information $I(A;f|t) = S(A|t) - S(A|f,t)$ for distributions conditional on both age and the FI. This is the information gained by knowing a specific range of the FI, as indicated in the legend, vs just knowing their age.  The negative values of $I(A;f|t)$ for older individuals with low frailties indicates that they have wider (normalized) death-age distributions compared to the population average at that age.}
\label{ModelInformationSpecificFI}
\end{figure}

Fig.~\ref{ModelInformationSpecificFI} shows the specific mutual information $I(A;f|t) = S(A|t) - S(A|f,t)$ vs age, which is the information gained by including a FI value in the given range at a given age, compared to just knowing their age. It is important to note that this is {\em not} comparing the predictive value of just the FI to the predictive value of just age, but rather the additional information provided by the FI while also knowing age. This specific mutual information is not averaged over all FI values, so it can be negative.  The negative values of $I(A;f|t)$ for older individuals with low frailties indicates that they have wider (normalized) death-age distributions compared to the population average at that age.  A larger FI is most informative for younger individuals --- and can exceed the information gained from knowing age alone. As age increases, the information along each specific FI curve decreases. This is due to the continually increasing average FI of the population, together with the narrowing of the death-age distribution due to increasing age $t$. 

\begin{figure}  
  \centering
  \includegraphics[width=0.46\textwidth]{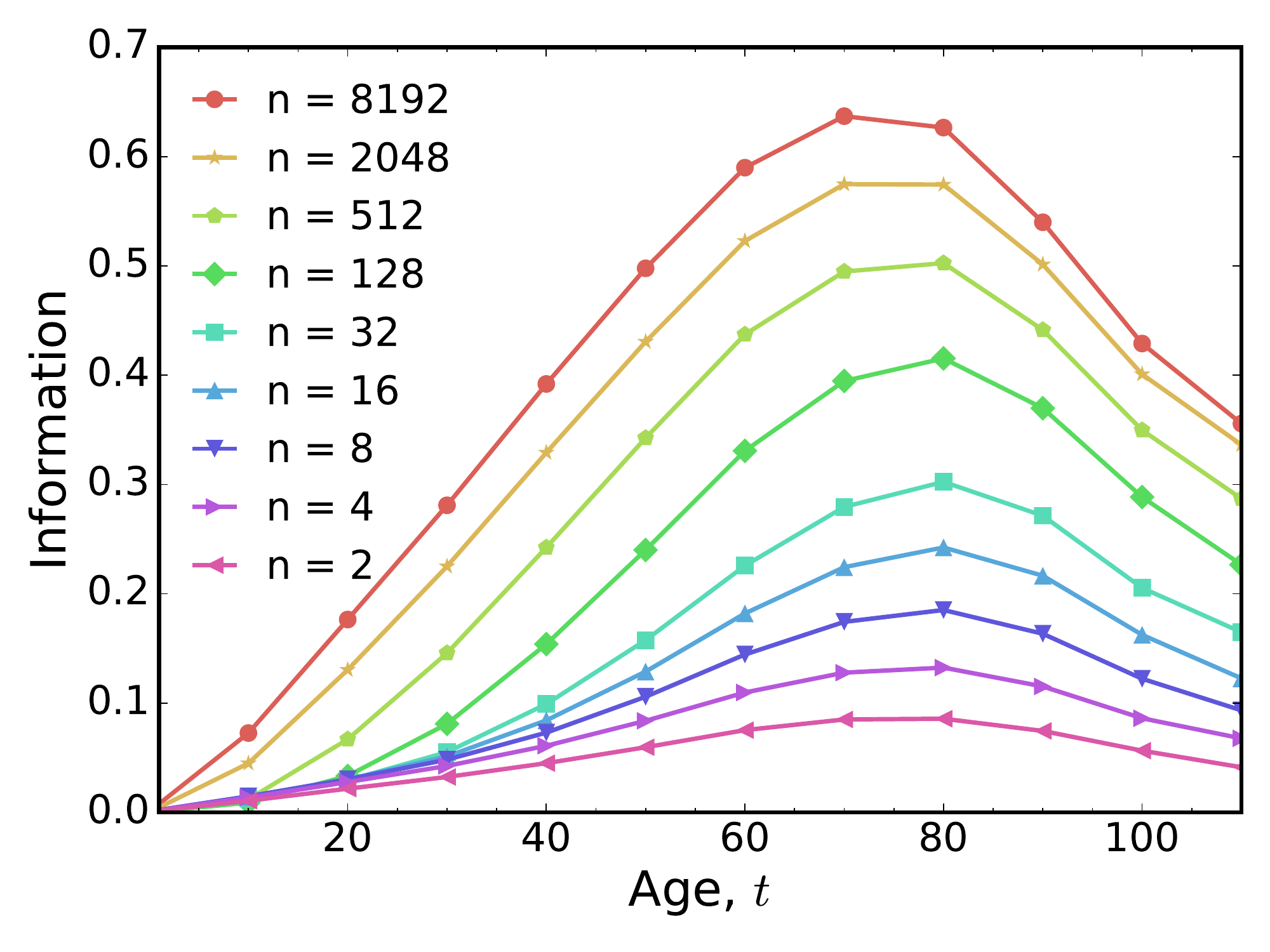}
  \caption{Mutual information conditioned on age $I(A;F|t)$ vs age. As indicated by the legend, the information increases with increasing number of deficits $n$ included in the FI. Otherwise, default model parameters were used -- including $q=0.3$.}
\label{MutualInformationVsAge}
\end{figure}

Fig.~\ref{MutualInformationVsAge} shows the value of the mutual information $I(A;F|t)$ for different numbers of deficits $n$, conditioned at different ages $t$. In contrast to Fig.~\ref{ModelInformationSpecificFI}, this information is averaged over all of the FI values. The peak around age $80$ means this is where the FI is most predictive on average. The decrease in information towards the youngest  ages is the result of the the preponderance of low FI in the population. For older individuals  age alone becomes very informative (see Fig.~\ref{ModelInformationAge})--- which reduces the additional information that can be provided by the FI.  As we increase the number of deficits included in the FI by constant factors we monotonically increase (approximately logarithmically) the predictive value of the FI.  

\begin{figure} 
  \centering
  \includegraphics[width=0.46\textwidth]{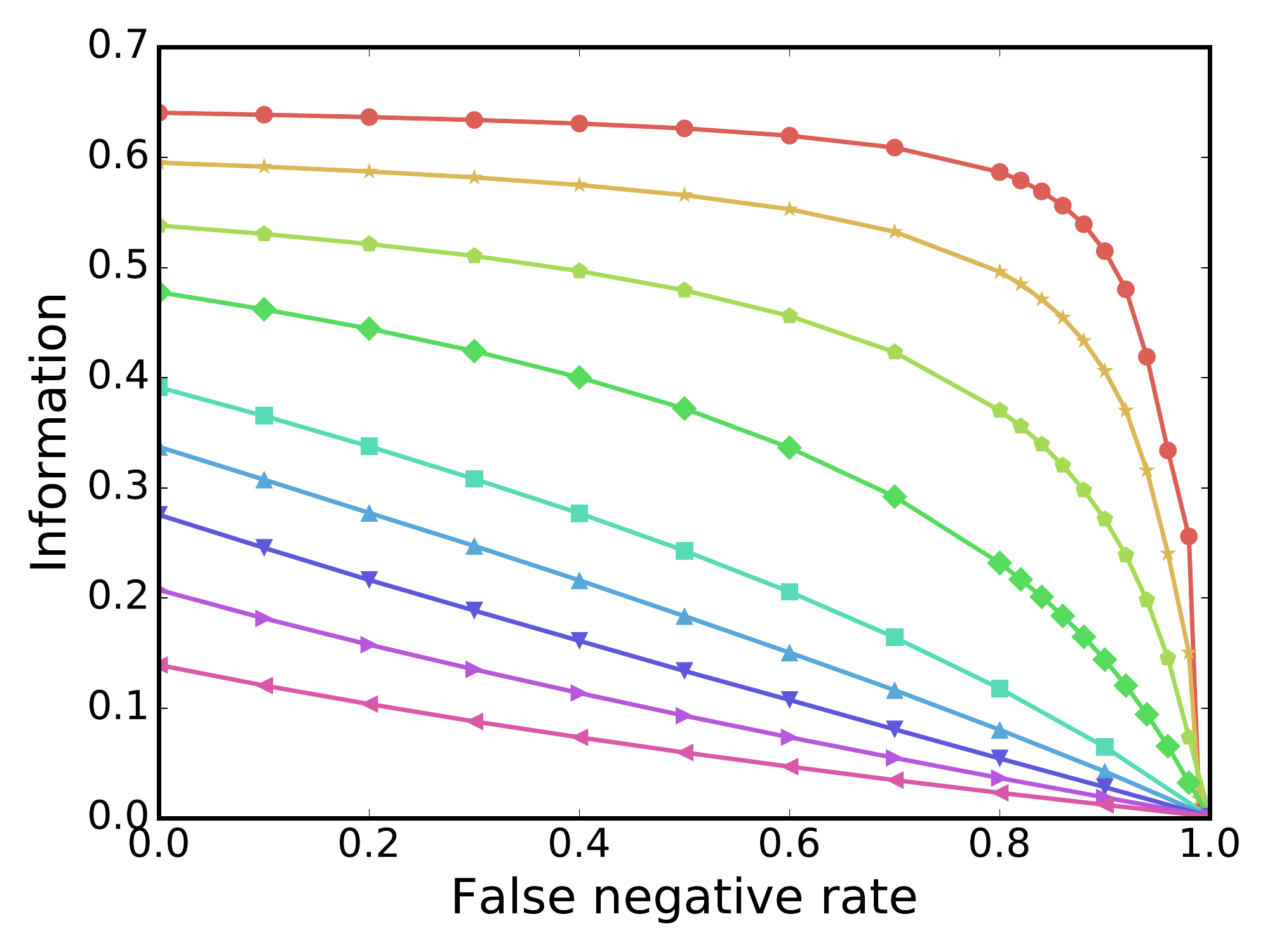}
  \caption{Mutual information at age $80$ years, $I(A;F |t = 80)$, vs the false negative rate $q$. Each curve corresponds to a different number $n$ of deficits in the FI, as indicated by the legend in Fig.~\ref{MutualInformationVsAge}. Other model parameters have default values.}
\label{InformationErrorRate}
\end{figure}

Fig.~\ref{InformationErrorRate} shows the effect of the false-negative rate $q$ on the mutual information provided by the FI, at age $t=80$ years (close to the peak from Fig.~\ref{MutualInformationVsAge}). As we expect, the average information provided by the FI decreases monotonically as  $q$ increases, and vanishes when $q=1$. However, for our default value of $q = 0.3$ there is only a modest decrease in the amount of information. We also see that increasing the number of deficits $n$ in the FI can offset the degradation due to $q$. For very large $n$, there is very little information loss until very large $q$. This is essentially because for large $n$ the false-negative rate still changes $f$ but no longer introduces significant stochasticity.

\begin{figure}  
    \includegraphics[width=.46\textwidth]{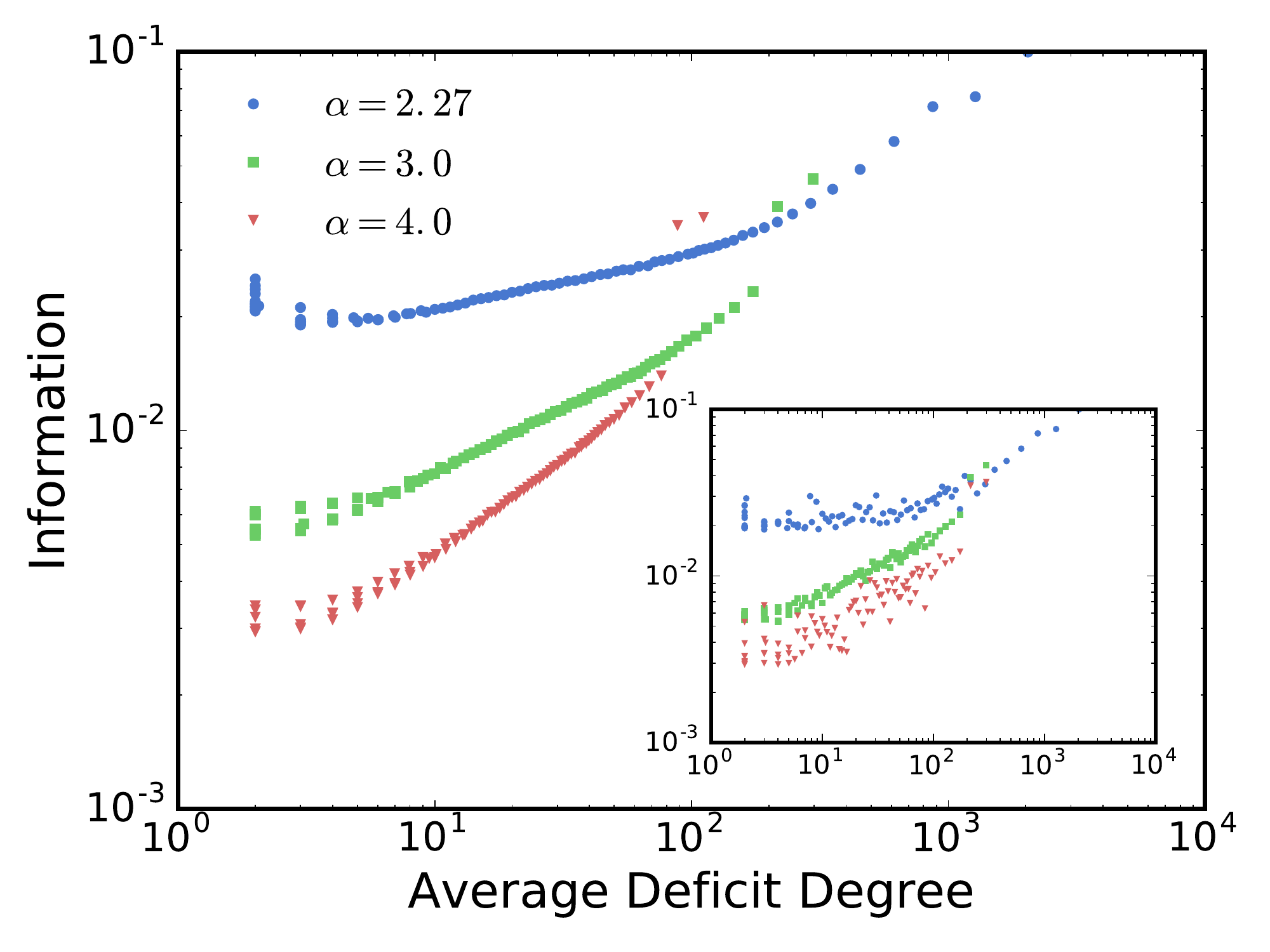}
    \caption{The information spectrum of our deficits: mutual information per deficit at age $80$, $I(A;D_i|t = 80)$ vs the average network degree of the deficit $\langle k_i \rangle$. Deficit indices are ranked in order of connectivity, and $i$ corresponds to deficits of the same order for different individuals. Different parameter values of the scale-free network exponent $\alpha$ are shown, as indicated by the legend. Other model parameters have default values. The main plot shows the simulation with a population of $10^7$, and the inset shows a population of $10^4$.}
    \label{ModelConnectionSpectrum}
\end{figure}

Mutual information allows us to reach into the network topology of our model. Fig.~\ref{ModelConnectionSpectrum} shows the information per deficit vs the average degree of these deficits; we call this the information spectrum of our model. The two highest degree points for each curve are the mortality nodes. These nodes do not follow the general trend on their respective curves, due to their unique role in the network. We see that normal deficits with a larger average degree tend to provide more information, with  an approximately power-law relationship at intermediate degrees.  These plots qualitatively explain the diminishing returns in information as more deficits are added to the FI in Figs.~\ref{MutualInformationVsAge} and \ref{InformationErrorRate}. Information is plotted for different values of the scale free network exponent, $\alpha$. The information spectrum gets steeper as $\alpha$ increases. Since the network degree distribution  also gets steeper, there are very few highly informative nodes at larger $\alpha$.  The inset shows the same analysis with a simulated population of $10^4$ individuals. We found that the information spectrum started to be reliable for populations of more than $10^3$ model individuals.
 
\section{Discussion} 

Our model is able to recover the average FI vs age, the exponential increase in Gompertz law of mortality rates, and the increasing variation in individual health through the broadening of the FI distributions with age. With our third hypothesis for $f_{max}$, the addition of a false-negative rate $q$, we could also recover  observed $f_{max}$ values.  By assuming that $q$ varies between studies, we naturally explain the observed non-universality of $f_{max}$ \cite{Drubbel:2013, Harttgen:2013, Clegg:2016, Searle:2008, Gu:2009, Mitnitski:2013, Bennett:2013, Hubbard:2015, Armstrong:2015}.

Like Taneja {\em et al} \cite{Taneja:2016}, we could not make the first hypothesis, that parameter tuning can recover $f_{max}$, work while retaining the Gompertz law and the average increase of FI with age -- despite much improved computational efficiency and the ability to vary the scale-free exponent $\alpha$.  Similarly, using an auxiliary mortality condition at $f_{max}$ to force the FI limit led to unobserved discontinuities in the distribution of FI at later ages (see Fig.~\ref{FIDistributions}(c)).   Even if they were made to work, these first two hypotheses would also need to invoke intrinsic differences in the aging and mortality processes between cohorts to explain the observed non-universality of $f_{max}$.

Binarized deficits, such as used in our model, require well-defined thresholds or cut-points between states~\cite{Searle:2008}. For example, continuous-valued blood biomarkers use thresholds to classify deficits~\cite{Mitnitski:2015}. For realistic measures, this binary classification introduces false positives and/or false negatives. This is a well-studied issue when dealing with binary classifiers of continuous measures~\cite{Zweig:1993}. A similar issue should arise with ordinal deficits, where there are multiple ranked levels of damage associated with the deficit~\cite{Searle:2008}. We note that such classification errors are reproducible, and do not represent avoidable noise or measurement error. Measurement errors would also contribute to false positives and false negatives~\cite{Dent:2015, Forti:2012, Pijpers:2012, Clegg:2015} but are, in principle, both random and correctable. Nevertheless, the false-negative rate $q$ in our model analysis does not distinguish between systematic classification errors and stochastic measurement errors. 

Typically, thresholds used to binarize deficits are determined by standard diagnostic criteria \cite{Clegg:2016} or empirically from population survival curves \cite{Mitnitski:2015}. As a thought-experiment, it is helpful to consider shifting every threshold (or cut-point) from their standard values. For large-enough thresholds, all deficits will always be classified as healthy and we will have $f_{max}=0$. In this limit, the sensitivity vanishes. For small-enough thresholds, all deficits will always be classified as damaged and we will have $f_{max}=1$. In this limit, the specificity (one minus the false-positive rate) vanishes. In between, we expect $f_{max}$ to continuously depend on the choice of thresholds.  The observation of $0<f_{max}<1$ necessarily follows from having both non-zero specificity and sensitivity. Our bare model deficits are idealized in this respect, since deficit damage perfectly correlates with increased local damage rates (perfect sensitivity) and healthy deficits never contribute to local damage rates (perfect specificity). Imposing $q>0$ on our model FI appears reasonable, and by doing it we impose a finite sensitivity with respect to further damage and mortality.  

False-negative errors, which correspond to limited sensitivity, are intrinsic to clinical assessment due to the tradeoff between specificity and sensitivity \cite{Metz:1978, Zweig:1993}. Sensitivity equals $1-q$. For age-related clinical measures, sensitivities of $ \approx 0.6-1.0$ are reported with respect to various mortality outcomes \cite{Clegg:2015} -- consistent with our overall $q=0.3$.   Similar sensitivities of clinical diagnostics are reported in internal medicine with respect to post-mortem autopsy results \cite{Anderson:1989}.  

Our current computational model, parameterized with a false negative rate, captures the aging phenomenology and appears reasonable. However, other mechanisms for $f_{max}<1$ might also contribute. We have included a fairly generic Barab\'{a}si-Albert scale-free network topology in our model. We have not explored more structured network topologies \cite{Watts:1998}, some of which can coexist with a scale-free degree distribution \cite{Ravasz:2003}.  Recent observational studies have distinguished between subclinical deficits (from e.g. blood tests, vital signs, or electrocardiographic measures) \cite{Blodgett:2016, Rockwood:2015, Mitnitski:2015, Howlett:2014} and clinical ones (from e.g. a comprehensive geriatric assessment, or CGA).  We can imagine that such classes of deficits evolve with different  parameters, or  differently with mortality or frailty deficits, and that this might allow $f_{max}$ to be tuned with model parameters.  

We use our efficient computational model (with $q=0.3$) to generate death age distributions of a large simulated population. Conditioning the population on the current age and/or current FI generally reduces the range of possible death ages. The effect of knowing a persons FI can be seen in the narrowing death age distributions at a given age and FI. This leads to an increase in the information known about a persons death age. With a narrower death age distribution, better estimates of life expectancy can be made. We quantify this increase in the predictive value with the mutual information. Mutual information is a non-parametric measure of the predictive value of the FI.  We  also use mutual information to begin to characterize the spectrum of information of individual deficits, and how they relate to local network topology.

The mutual information $I(A;F|t)$ gives us a way of measuring the average reduction in uncertainty in the death age, at a given age,  by knowing the FI. The information shows how well the FI correlates with the death age. It is a measure how well the FI can be used as a proxy of health, with respect to mortality. We find that this value has a maximum at around $80$ years old. This means that on average, the FI will be most informative of a persons death age when the person is around age $80$. As age increases from $80$, people die with both large and small FI values, making the FI less informative. Similarity for ages much smaller than $80$, most people have a low and uninformative FI. 

The specific mutual information $I(A;f|t)$ gives us the predictive value of a specific range of FI values. The FI is most predictive at large values. Age is always a strong factor in how predictive the FI is, as was seen with, e.g., individual risk factors of heart disease~\cite{Blokh:2015}. This is because the predictive value of the FI depends on differences between the conditioned subpopulation and the general population. If a large proportion of the population have the same FI, this value of the FI does not offer much in addition to just knowing their age.  Even at very low values of the FI, age itself eventually becomes more predictive of the death age than the FI. As can be seen in Fig.~\ref{ConditionalDistributions}, death occurs much later for younger individuals with low FI than for much older individuals with the same FI. We see similar results in population data (see, e.g., Fig.~2 of \cite{Clegg:2016}).  This is the result of the FI not encapsulating the full extent of damage in an individual, even though model mortality is only due to accumulative damage.

The information content of the FI decreases with an increasing false negative rate $q$. However, we see only a small decrease for the false negative rate of $0.3$ used in the model to recover the FI limit.  Balancing this, the information content of the FI increases as the number of deficits included increases. Qualitatively, a deficit spectrum suggests that including large numbers of deficits in the FI will lead to diminishing returns. Indeed, Fig.~\ref{InformationErrorRate}, shows that the information increases approximately logarithmically as the number of deficits increases. Nevertheless, our model parameterization does not show any evidence that large numbers of deficits dilutes or diminishes the predictive value of the FI. This is in qualitative agreement with observational data \cite{Song:2014, Rockwood:2007}.

We have shown that the information spectrum of deficits, shown in Fig.~\ref{ModelConnectionSpectrum}, is strongly dependent on the network topology through the scale-free exponent $\alpha$ --- with an approximately power-law dependence. We also found that (see Appendix~B) $\alpha$ strongly affects mortality statistics.  Reinforcing this, deficits in a deterministic model without network structure (see Appendix~A) significantly changes the mortality behavior of the model, as well as the evolution of the FI.  Probing the network structure of age-related deficits will be desirable to estimate $\alpha$ and $\langle k \rangle$ directly.

Interestingly, our model parameterization shows little sensitivity to deficit repair rate (through $R$ or $\gamma_-$, see Appendix~B). For our model, this is because damage rates are so strongly affected by local frailty through $\Gamma_+(f)$. Effectively, most damage occurs when the local frailty is substantial and so any repair is soon redamaged. Again, for our model, this implies that deficit repair does not affect longevity statistics or the overall FI. It will be interesting, and important, to assess the rate and significance of deficit repair in clinical populations. To do this, we hope to undertake further analysis of longitudinal studies in which frailty-trajectories (individual time-series) are recorded.  Since a thorough exploration of parameter space is not possible due to the ``curse of dimensionality'',  such direct estimation of model parameters from observational data is also needed to test or identify the `correct' parameterization of our model for human mortality studies.

Our model allows us to rapidly generate large quantities of high-quality data. For our model, information measures appear to be useful and reliable with cohort sizes in excess of $\approx 10^3$ individuals --- which is towards the largest of traditional observational cohorts. Large quantities of clinical health data with over $10^5$ individuals are now becoming available through electronic health records~\cite{Clegg:2016}. We have used information measures with our model data as a first step towards applying them to these emerging electronic records. We believe that non-parametric information measures will be an important tool for characterizing data-sets of large cohorts, and will lead to greater understanding of the relationships between mortality and health deficits. 

\begin{acknowledgements}

We thank ACENET for computational resources, along with a summer fellowship for SF.  ADR thanks Natural Sciences and Engineering Research Council of Canada (NSERC) for operating grant RGPIN-2014-06245. We thank Dr. Danan Gu for providing us the population data for Fig.~\ref{FIDistributions} \cite{Gu:2009}. 

\end{acknowledgements}

\appendix   
\section{Deterministic network model}
\label{appendix:deterministic}

In this appendix we present a deterministic ``mean-field'' model of aging that captures some of the basic phenomenology, but treats all deficit nodes identically. Formally, we consider a maximally connected network in which all nodes are connected to all other nodes. For computational convenience, we also take the limit as the number of deficits $N \to \infty$ and as the number of FI deficits $n \to \infty$. This also demonstrates that those limits are well behaved. We can then write rate equations for the dynamical processes, since every deficit will have the same local frailty $f$ that is identical with the global frailty.  
   
The FI evolves as 
\begin{equation} \label{fequation}
	\dot{f}(t) = (1-f)\Gamma_{+}(f) - f\Gamma_{-}(f),
\end{equation}
where, as before, $\Gamma_+= \Gamma_0 e^{\gamma_+ f}$ and $\Gamma_- = (\Gamma_0/R) e^{- \gamma_- f}$.
Mortality is determined by separating the population into subpopulations, dependent on the state of their mortality nodes (we consider two mortality nodes, as in the full computational model, but this mean field approach can be adapted to include any number of mortality nodes). Let $N_{0}$ be the proportion of people with two healthy mortality nodes, $N_1$ be the proportion with one damaged mortality node, and $N_{2}$ be those with two damaged mortality nodes (i.e. those that are deceased by our mortality rules).  Transitions between these subpopulations occur by damaging or repairing mortality nodes, so that we obtain simple dynamics
\begin{eqnarray}
	\dot{N}_{0}(t) &=& \Gamma_{-}(f) N_1 - 2\Gamma_{+}(f) N_{0} \\
	\dot{N}_{1}(t) &=& 2\Gamma_{+}(f) N_{0} - N_{1} (\Gamma_{+}(f) + \Gamma_{-}(f)) \\
	\dot{N}_{2}(t) &=& \Gamma_{+}(f) N_{1}.
\end{eqnarray}
Initially we take $N_0(0)=1$ with $f(0)=0$, corresponding to the initial conditions of our full network model.
We can check that $N_0+N_1+N_2=1$.  The current alive fraction will be $N(t) \equiv N_{1}(t) + N_{0}(t)$, and the current deceased population $N_{2}(t)$. The instantaneous mortality rate is given by $\mu(t) = \dot{N}_{2}(t)/N(t)$. We note that since all nodes are  connected to all others, $f$ is not a stochastic variable (i.e. the distribution of $f$ is a delta-function).  Therefore age and FI provide the same information about death-ages, and we have no mutual information with FI in the mean-field model, i.e. $I(A;f|t)=I(A;F|t)= I(A;D_i|t)=0$.

Our ``mean-field'' model is deterministic. Furthermore, we obtain the {\em same} dynamical equations if we impose  the same deterministic evolution Eqn.~\ref{fequation} on each local frailty $f_i$ of the $i$th node, since the only symmetry-breaking mechanism between nodes is stochastic.  The network topology is only significant in a stochastic model.

\section{Parameter dependence} 
\label{appendix:parameters}

\begin{figure} 
  \centering
  \begin{minipage}{0.46\textwidth}
    \includegraphics[width=\textwidth]{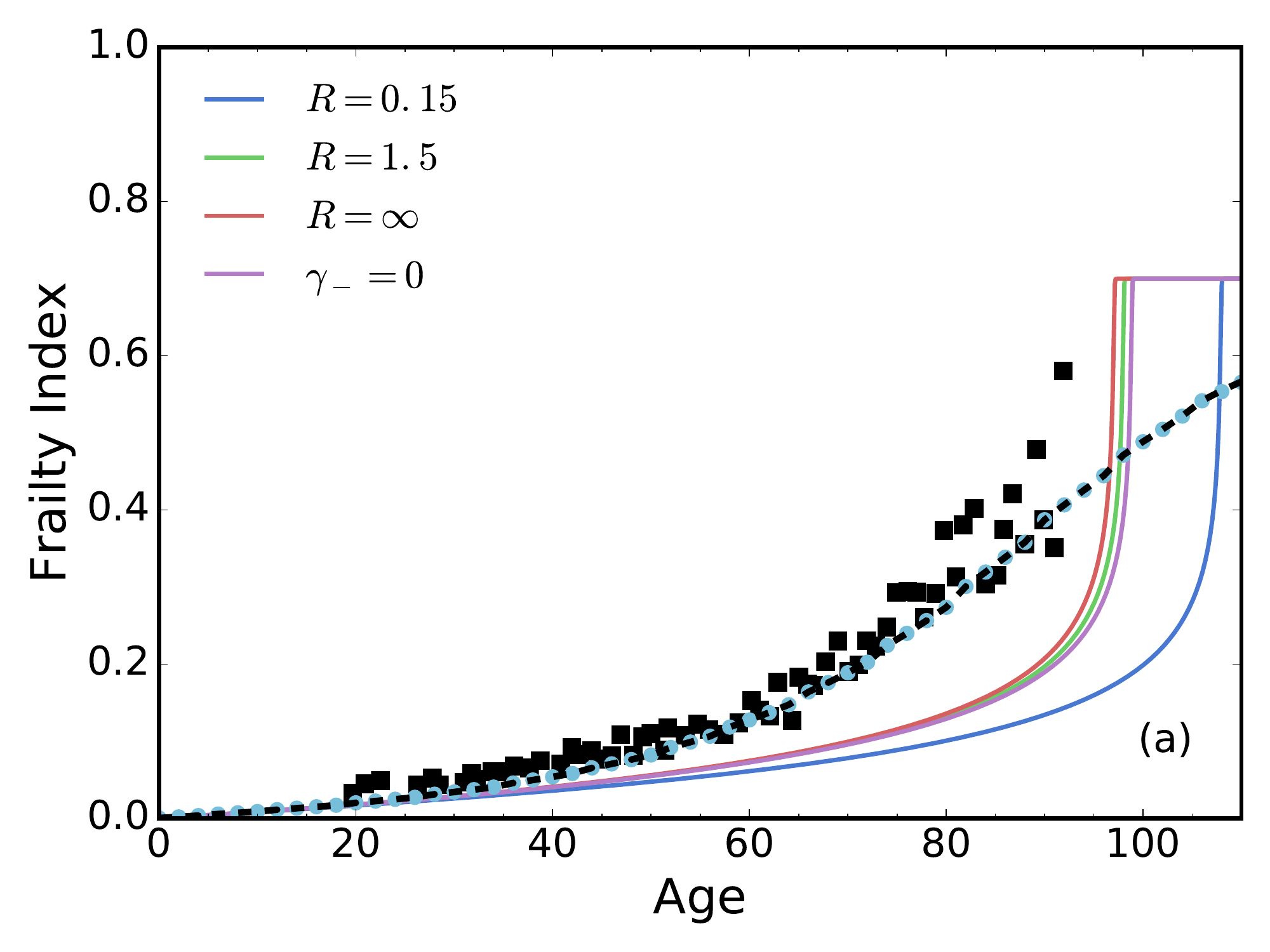}
        \includegraphics[width=\textwidth]{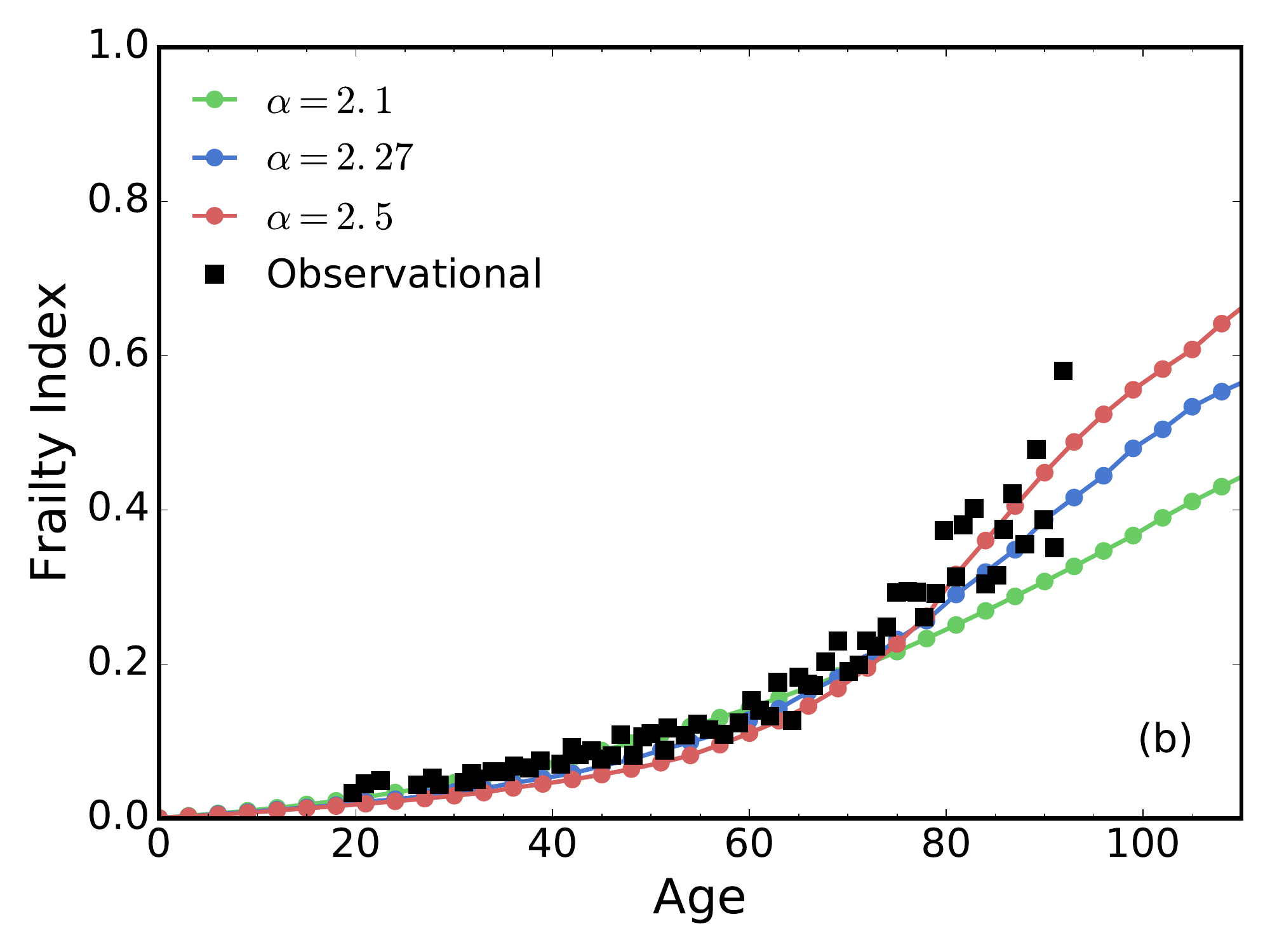}
    \caption{FI vs age, using default model parameters except as indicated in the legends. The black points are the same population data shown in Fig.~\ref{Frailtyq03}.  (a) Solid lines indicate our deterministic model from Appendix~A. The red line ($R=\infty$) has repair rates turned off, while the purple line ($\gamma_-=0$) has the suppression of repair rates by local frailty turned off. In both cases, the results are close to the default parameters (green line, $R=1.5$). Only when the initial repair rate greatly exceeds the initial damage rate (blue line, with $R=0.15$) does the FI begin to grow more slowly with age. The light blue points are the same network model data shown in Fig.~\ref{Frailtyq03}, while the dashed black line superimposing the light blue points are network model data with repair turned off ($R = \infty$). (b) The network scale-free exponent $\alpha$ is varied as indicated.}
    \label{MFTFrailty}
  \end{minipage}
  \end{figure}

  \begin{figure}	
  \begin{minipage}{0.46\textwidth}
    \includegraphics[width=\textwidth]{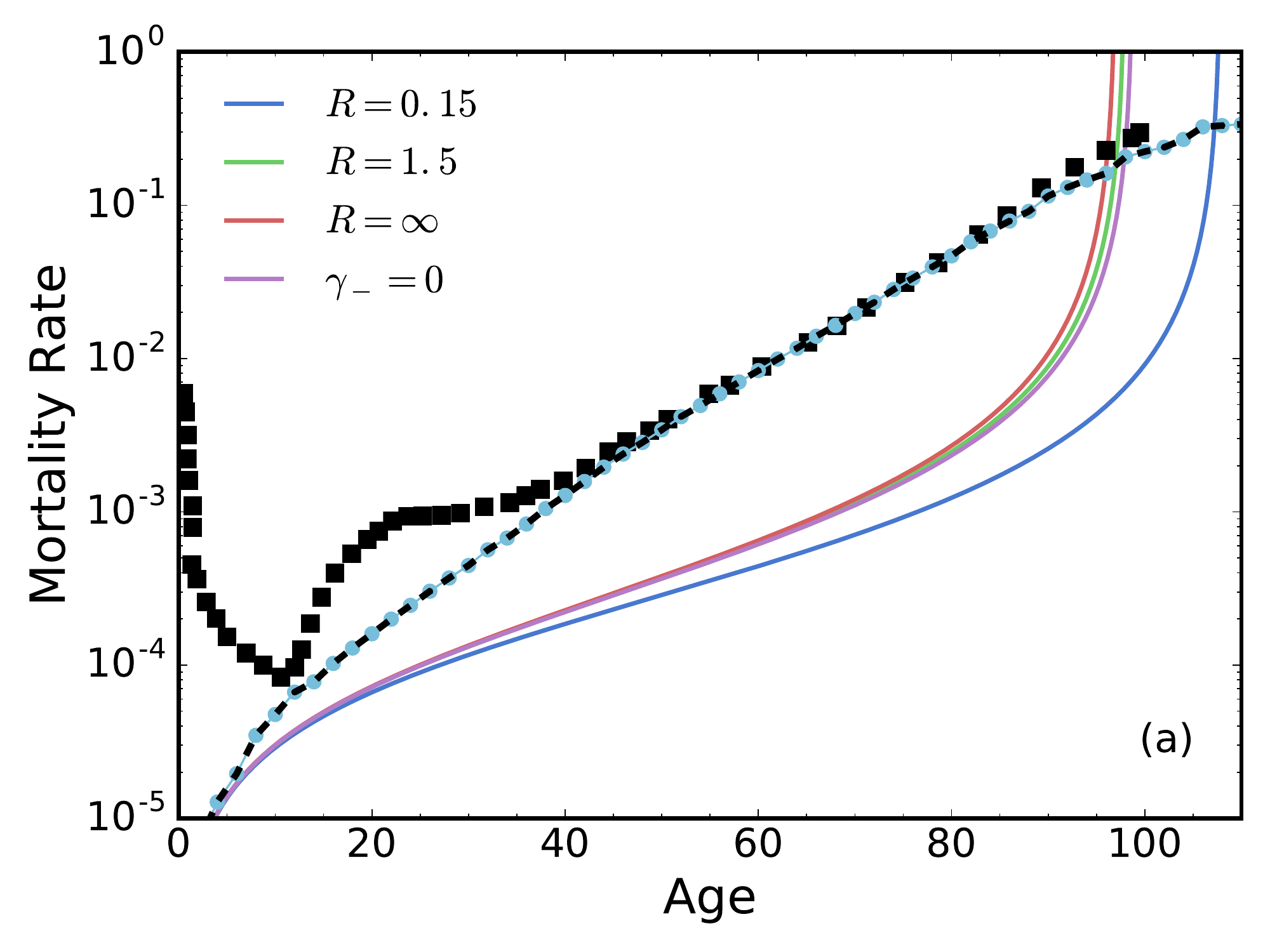}
        \includegraphics[width=\textwidth]{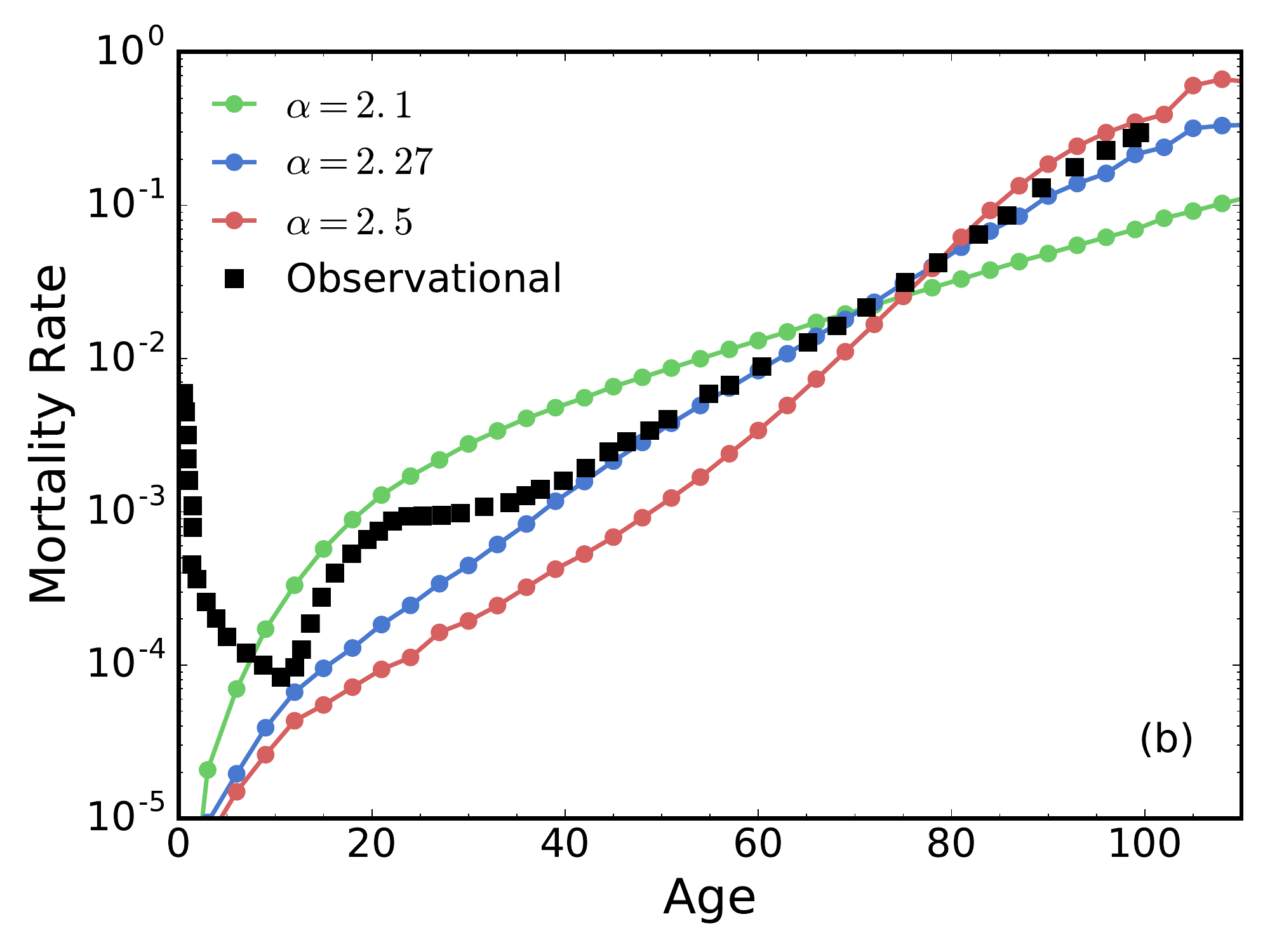}
    \caption{Mortality rate vs age, using  default parameters, except as indicated in the legend. The black points are the same population data shown in Fig.~\ref{Mortality}. (a) Solid lines indicate our deterministic model from Appendix~A.  The green line ($R = 1.5$) has identical parameters as the network model. The light blue points are the same network model data shown in Fig.~\ref{Mortality}, while the dashed black line superimposing the light blue points are network model data with repair turned off ($R = \infty$). (b) The network scale-free exponent $\alpha$ is varied as indicated.}
    \label{MFTMortalityRate}
  \end{minipage}
\end{figure}

Fig.~\ref{MFTFrailty} (a) shows the FI vs age for our deterministic model. We have used our default parameterization (with $q=0.3$), except where indicated by the legend. The false negative rate is applied by multiplying $f$ by $1-q$. We have slower growth of $f$ vs $t$, but then rapid growth towards $f_{max} \approx 1-q$.  As indicated by the legend, we can vary repair significantly and not qualitatively change $f(t)$ in our deterministic model. This is also seen in our full network model with the agreement between default parameters (blue circles) with repair turned off ($R=\infty$, dashed black line).  Repair appears not to be an important process for our model, for our default parameterization.  In (b) we see that the scale-free network exponent $\alpha$ affects the evolution of the FI at later ages.

Fig.~\ref{MFTMortalityRate} (a) shows the mortality rate vs age for our deterministic model. We have used our default parameterization, except where indicated by the legend. The data from our full network model (light blue points) agrees only at the youngest ages. At later ages, our deterministic model significantly underestimates mortality.   The network topology allows our full computational model to much better capture the aging phenomenology.  Again, turning repair off (red line with $R=\infty$) does not significantly change the mean-field results. As shown by the dashed black line, turning repair off does not change the mortality of our full network model. We are in a parameter regime of the  model where repair is not significant for mortality statistics or for the evolution of the FI. 

Interestingly, Fig.~\ref{MFTMortalityRate} (b) indicates that the scale-free network exponent $\alpha$ strongly affects mortality statistics. This is in significant contrast with the relative independence of mortality on network parameters reported in earlier studies \cite{Vural:2014, Taneja:2016}. However, those studies did not vary $\alpha$. This $\alpha$ dependence emphasizes the need to characterize network topology in observational studies, with e.g. the information spectrum of Fig.~\ref{ModelConnectionSpectrum}.

\bibliography{ref}
\end{document}